\documentclass[12pt,preprint]{aastex}

\shorttitle{Leptonic Model of Steady $\gamma$-Ray Emission from Sgr A$^*$}
\shortauthors{Kusunose, \& Takahara}

\begin{document}

\title{A Leptonic Model of Steady High-Energy Gamma-Ray Emission 
from Sgr A$^*$}

\author{Masaaki Kusunose}
\email{kusunose@kwansei.ac.jp}
\affil{Department of Physics, School of Science and Technology,
Kwansei Gakuin University, Sanda 669-1337, Japan}

\and

\author{Fumio Takahara}
\email{takahara@vega.ess.sci.osaka-u.ac.jp}
\affil{Department of Earth and Space Science,
Graduate School of Science, Osaka University,
Toyonaka 560-0043, Japan}

\begin{abstract}
Recent observations of Sgr A$^*$ by {\it Fermi} and HESS have detected
steady $\gamma$-ray emission in the GeV and TeV bands.
We present a new model to explain the GeV $\gamma$-ray emission
by inverse Compton scattering by nonthermal electrons supplied by
the NIR/X-ray flares of Sgr A$^*$.
The escaping electrons from the flare regions accumulate in a region with
a size of $\sim 10^{18}$ cm and magnetic fields of $\lesssim 10^{-4}$ G.
Those electrons produce $\gamma$-rays by inverse Compton scattering off
soft photons emitted by stars and dust around the central black hole.
By fitting the GeV spectrum, we find constraints on the magnetic field
and the energy density of optical-UV radiation in the central 1 pc region around
the supermassive black hole.
While the GeV spectrum is well fitted by our model,
the TeV $\gamma$-rays, whose spectral index is different from 
that of the GeV emission, may be from different sources such as 
pulsar wind nebulae.
\end{abstract}

\keywords{
black hole physics
--- Galaxy: center
--- plasmas
--- radiation mechanisms: non-thermal
}

\section{Introduction} \label{sec:intro}

Sagittarius A$^*$ (Sgr A$^*$) is located at the center of our Galaxy
and harbors a massive black hole
\citep[see][for review]{mel07,gen10}.
Recent observations have shown that the distance to Sgr A$^*$ is
$\sim 8$ kpc \citep{eis03}
and the black hole mass is $\sim 4 \times 10^6 M_\sun$ 
\citep[e.g.,][]{ghez09,gill09a,gill09b}.
The bolometric luminosity of Sgr A$^*$,
$L_\mathrm{bol} \sim 10^{36}$ erg s$^{-1}$, is dominated by radio
and the peak in the $\nu F_\nu$ representation occurs at $\sim 10^{12}$ Hz
\citep{zyl95,fal98}.
In the quiescent state, X-ray (2 -- 10keV) luminosity is very dim, 
i.e., $L_X \sim 2.4 \times 10^{33}$ erg s$^{-1}$ \citep{de09,de11}.
However, frequent flares are observed in the X-ray band as well as
the near infrared (NIR) band \citep{de09,de11}.
In the high energy regime, TeV $\gamma$-rays have been observed
by CANGAROO \citep{tsu04}, VERITAS \citep{ko04}, HESS 
\citep{rh05,aha04,aha06,aha08,aha09},
and also by MAGIC \citep{albert06}.
The HESS source is named HESS H1745--290.
Recent observations show that 
TeV emission exhibits no time variation \citep{rh05,albert06,aha08}.
More recent observations by {\it Fermi} Large Area Telescope 
find that GeV $\gamma$-rays are emitted in the region coinciding with
Sgr A$^*$ \citep{abdo09,coh09}. 
The source is named 1FGL J1745.6--2900.
The observed GeV $\gamma$-rays $> 300$ MeV are from 
the region around Sgr A$^*$.
The averaged flux of GeV $\gamma$-rays is $(324.9 \pm 7.05) \times 10^{-9}$
counts cm$^{-2}$ s$^{-1}$ and there is no statistically significant 
variability.
The spectrum is well fitted by a broken power law with the break energy
$E_\mathrm{br} = 2.0^{+0.8}_{-1.0}$ GeV
and the power law indices $\Gamma_1 = 2.20 \pm 0.04$ ($E < E_\mathrm{br}$)
and $\Gamma_2 = 2.68 \pm 0.05$ ($E > E_\mathrm{br}$)
\citep{cher11}.

While the emission in radio through infrared and possibly X-ray bands is
explained by emission from radiatively inefficient accretion
flow (RIAF) \citep{yuan03} or jets \citep{fm00}
near the central black hole,
the emission mechanisms of high-energy radiation are still debated.
Before the launch of {\it Fermi}, 
\cite{ad04} proposed a MHD wind shock model for the TeV emission from Sgr A$^*$.
The TeV emission observed by HESS was modeled also by \cite{bal07} 
(hadronic model) and \cite{ha07} (leptonic model).
In the model by \cite{ad04}, electrons with $\gamma \lesssim 10^8$
scatter photons with $\nu \sim 10^{12}$ Hz from RIAF and 
far-infrared dust radiation.
This produces TeV $\gamma$-rays in the Thomson scattering regime.
\cite{ha07} show that TeV emission is explained by inverse Compton 
(IC) scattering off infrared (IR) and optical photons 
in pulsar wind nebula G359.95--0.04.
The flux of GeV $\gamma$-rays of these models is
smaller than the flux observed by {\it Fermi} recently.
The contribution of high energy sources in the Galactic
center region is reviewed by \cite{cja11}.
Recently, \cite{cher11} proposed a hadronic model to explain the spectrum
both in the TeV and GeV bands,
assuming that the sources of HESS and {\it Fermi} are coincident.
Relativistic protons injected by Sgr A$^*$ 
interact with ambient matter and produces $\gamma$-rays.
For example, a constant injection of high-energy protons for 
$10^4$ years reproduces the observed very high energy
$\gamma$-ray spectrum.
The different spectral shapes in the GeV and TeV bands
are owing to the different effective speeds of the protons 
through the ambient matter.

Although the attenuation of TeV photons by $e^+ e^-$ pair production
may change the spectral shape of the TeV $\gamma$-rays,
this is not the case for Sgr A$^*$.
As shown by several authors  \citep{ps05,mps06,zhan06},
the attenuation of TeV photons by $e^+ e^-$ production on
the Galactic interstellar radiation field is weak
for photon energy less than 10 TeV \citep[see also][]{an05}.

Because HESS and {\it Fermi} do not have enough spatial resolution,
the coincidence of both sources, HESS H1745 -- 290 and 1FGL J1745.6 -- 2900, 
is not conclusive.
In this paper we present another model of the steady $\gamma$-ray 
emission, focusing on the GeV emission.
We recently proposed a synchrotron blob model to explain the NIR/X-ray
flares from Sgr A$^*$ \citep{kt11}.
In this model the temporal injection of electrons is assumed to produce 
flares by synchrotron radiation.
The frequency of flare events is high,
e.g., the peaks of the light curves occur once a day and four times 
a day in X-ray and NIR bands, respectively \citep{bag03b,eck06,de11,trap11}.
Nonthermal electrons escape from the flare emission region on timescale
$\sim$ several $R/c$, 
where $R \sim 10^{13}$ cm is the size of the flare emission region
and $c$ is the speed of light.
Away from the flare region escaping electrons are accumulated 
owing to the ambient magnetic fields and
emit radiation through interaction with the magnetic fields 
and ambient radiation fields emitted by stars and dust.
There, away from the central accretion flow,
the strength of magnetic field is smaller than that in the accretion flow, 
and IC scattering becomes a dominant radiative process.
The Lorentz factor of nonthermal electrons of the flare model is
about $10^4$ and the ambient radiation field has a peak at $\nu \sim 10^{15}$ Hz
\citep{mez96} in a region $< 1.2$ pc.
Then it is expected that photons with $\nu \sim 10^{23}$ Hz are produced 
by IC scattering and this is in the GeV band observed by {\it Fermi}.
In this paper we show numerically that the emission by IC scattering naturally 
explains the GeV emission from Sgr A$^*$.

We describe our model in Section \ref{sec:model} 
and show numerical results in Section \ref{sec:results}.
Finally, we discuss our results in Section \ref{sec:discussion}.


\section{Emission Model} \label{sec:model}

We assume that high energy electrons are supplied by the flare events 
that occur near the central black hole.
Although there are various high energy sources such as supernova remnants
in the Galactic center region, 
we assume that the injection of high-energy particles from the central region 
is dominant.
A spherical geometry with radius $r_\gamma$ is assumed for 
the high-energy (HE) $\gamma$-ray emission region.
We solve the kinetic equations of electrons and photons simultaneously
to obtain the spectra of electrons and photons self-consistently.
In the following we describe our model in detail.

In a steady state, nonthermal electrons are injected at rate $q_\mathrm{inj}(\gamma)$ 
per unit volume and unit interval of $\gamma$,
where $\gamma$ is the electron Lorentz factor,
and they escape from the HE emission region on timescale $t_\mathrm{esc}$.
The kinetic equation of the electrons in a steady state is given by
\begin{equation}
    -\frac{\partial}{\partial \gamma} [ \dot{\gamma}_\mathrm{rad}
    n_e(\gamma) ]  - \frac{n_e(\gamma)}{t_\mathrm{esc}} + q_\mathrm{inj}(\gamma) 
    = 0, 
\end{equation}
where $n_e(\gamma)$ is the electron number density per unit interval 
of $\gamma$ and $ m_e c^2 \dot{\gamma}_\mathrm{rad}$ $(<0)$ is 
the radiative cooling rate of an electron
with $m_e$ being the electron mass.
The emission mechanisms are synchrotron radiation and IC scattering.  
Here the soft photon sources are the synchrotron radiation by the
nonthermal electrons in the HE emission region 
(synchrotron self-Compton, or SSC)
and the photons emitted by stars and dust (external Compton scattering).
The average magnetic field of the HE emission region is denoted by $B$.
Since the magnetic field is weaker in the HE emission region 
than in the flare region with $\sim 20$ G, IC scattering is 
the dominant radiation process.
As shown in Section 3 below, the magnetic field $\lesssim 10^{-4}$ G and
the soft photon energy density $\sim 5 \times 10^4$ eV cm$^{-3}$ are found to be
typical values in the GeV emission region.
For these values of the magnetic field and soft photon energy density, 
the radiative cooling time of electrons is longer than 
$\sim 4 \times 10^{9}$ s for electrons with the Lorentz 
factor $\lesssim 10^5$.  Thus the cooling time is longer than the escape time
and the use of the steady-state homogeneous model is justified.
Here the escape time is set to be $20 r_\gamma/c$ in the numerical calculations.

In our flare model, we assumed the injection spectrum of electrons such as
\begin{equation}
    \label{eq:qin-flare}
    q_\mathrm{inj}^f(\gamma) = K_e^f \gamma^{-p} \exp(-\gamma/\gamma^f_\mathrm{max}) 
    H(\gamma - \gamma^f_\mathrm{min}) ,
\end{equation}
where $H(z)$ is the Heaviside function.
Here $K_e^f$, $p$, $\gamma^f_\mathrm{max}$, and $\gamma^f_\mathrm{min}$ are 
parameters.
Because the flare interval is a few hours and much shorter
than the dynamical timescale of the emission region 
$r_\gamma/c \sim 3 \times 10^7$ s,
we assume the continuous injection of electrons in our model.
By fitting the observations, we obtained $p = 1.3$ and 
$\gamma^f_\mathrm{max} = 5 \times 10^4$ (model A in \citet{kt11}).
The value of $\gamma^f_\mathrm{min}$ was 2.
We found that the electrons in the flare region rapidly cool 
and obey a broken power law approximately.
Because we assume that the electrons responsible for the steady HE 
$\gamma$-rays are supplied by the flares, 
we use a broken power-law spectrum of electrons as 
the injection spectrum of electrons into the HE emission region.
Namely,
\begin{equation}
    \label{eq:qin-HE}
    q_\mathrm{inj}(\gamma) = K_e [ 
    \gamma^{-p}  H(\gamma - \gamma_\mathrm{min}) 
    H(\gamma_\mathrm{br} - \gamma) 
    + \gamma^{-p_u}  H(\gamma - \gamma_\mathrm{br}) 
    H(\gamma_\mathrm{max} - \gamma) ] ,
\end{equation}
where $K_e$, $\gamma_\mathrm{min}$, $\gamma_\mathrm{br}$, 
$\gamma_\mathrm{max}$, and $p_u$ are parameters.
Note the difference of $\gamma^f_\mathrm{max}$ and $\gamma_\mathrm{max}$ in 
equations (\ref{eq:qin-flare}) and (\ref{eq:qin-HE}), respectively.
The value of $K_e$ is determined by the injection rate per unit volume, 
$q_\mathrm{inj, \, 0}$,
i.e.,
\begin{equation} 
    q_\mathrm{inj, \,0} = \int_{1}^{\infty}
    q_\mathrm{inj}(\gamma) d \gamma.
\end{equation}   
We set $\gamma_\mathrm{min} = 2$ and $p = 1.3$ as in a flare model
(model A in \citet{kt11}).
In the flare region we obtained $p_u \sim 2.54$, $\gamma_\mathrm{br} \sim 500$,
and $\gamma_\mathrm{max} \sim 2 \times 10^5$.
To fit the {\it Fermi} data, we use $p_u$ and $\gamma_\mathrm{max}$ as parameters,
while $\gamma_\mathrm{br}$ is set to be 500.
The value of $\gamma_\mathrm{br}$ is not important to fit the spectrum
in the GeV band.

In the Galactic center region soft photons are emitted by stars, dust,
and plasmas.
The radiation field in the central $30''$ ($\sim 1.2$ pc) region
is given in Figure 37 in \cite{mez96}.
In their figure, the emission in $\nu < 2 \times 10^{11}$ Hz is 
dominated by free-free emission,
dust emission dominates in 
$2 \times 10^{11} \mathrm{Hz} \lesssim \nu \lesssim 3 \times 10^{13}$ Hz,
stellar radiation in 
$3 \times 10^{13} \mathrm{Hz} \lesssim \nu \lesssim 2 \times 10^{16}$ Hz,
and hot plasmas emit X-rays in $\nu > 2 \times 10^{16}$ Hz.
From their figure the soft photon energy density $u_\mathrm{soft}$ is calculated
as $9 \times 10^{-7}$ erg cm$^{-3}$ or $6 \times 10^5$ eV cm$^{-3}$.
On the other hand, \cite{ha07} assumed the photon energy density
$5000$ eV cm$^{-3}$ both in optical-UV (3 eV) and NIR (0.3 eV)
as a radiation field model of the Galactic center (Table 1 in their paper).
They refer to the work by \cite{dav92} for the soft photon energy density.
Because there is uncertainty in the optical-UV energy density,
we assume that the photon spectrum in the optical-UV band
is approximated by a thermal radiation
with temperature $T_\mathrm{opt\mathchar`-uv}$ and energy density 
$u_\mathrm{opt\mathchar`-uv}$.
On the other hand, the IR spectrum is adopted from \cite{mez96}.
In Figure \ref{fig:softphoton-spec}, we show an example of the soft
photon spectrum used in our models.
Although X-ray emission is shown in \cite{mez96}, 
we do not include X-rays as soft photons,
because the Klein-Nishina effect suppresses the IC scattering of X-rays.

\section{Numerical Results} \label{sec:results}

Numerical calculations are performed with parameters such as
$p_u$, $\gamma_\mathrm{max}$, $q_\mathrm{inj, \, 0}$, $B$,
$T_\mathrm{opt\mathchar`-uv}$, and $u_\mathrm{opt\mathchar`-uv}$.
Other parameters are fixed: $r_\gamma = 10^{18}$ cm,
$t_\mathrm{esc} = 20 r_\gamma /c$, $\gamma_\mathrm{min} =2$,
$\gamma_\mathrm{br} = 500$, and $p = 1.3$.
Because the HE $\gamma$-ray emission is steady during {\it Fermi} observations,
the size of the emission region is greater than $\sim 1$ lt-yr and
we set $r_\gamma = 10^{18}$ cm.
The emission spectra by electrons are not much dependent on 
the value of $t_\mathrm{esc}$,
but the value of $q_\mathrm{inj, \, 0}$ is inversely proportional to $t_\mathrm{esc}$.

In Figure \ref{fig:sed-fit-tuv1ev}, 
spectral energy distributions (SEDs) are compared with the observed data.
Model parameters are given in Table \ref{table:param-values1}.
In this figure,
fixed parameters are $\gamma_\mathrm{max} = 1.7 \times 10^5$, 
$B = 10^{-4}$ G, and $T_\mathrm{opt\mathchar`-uv} = 1$ eV.
When $B = 10^{-4}$ G, the gyro-radius is $1.7 \times 10^{12}$ cm for $\gamma = 10^5$,
which is short enough to confine nonthermal electrons in the HE 
emission region by magnetic fields.
The values of $p_u$ and  $\gamma_\mathrm{max}$ are different from our flare model
for 2007 April 4, and this may be possible
because the values of $p_u$ and $\gamma_\mathrm{max}$ may be different from 
flare to flare.
In the spectrum, the emission below $\sim 10^{14}$ Hz is by synchrotron radiation.
Because the magnetic field is weak ($10^{-4}$ G), 
the flux by SSC component is negligible.
There are two breaks in the SEDs.
Namely, breaks at $\nu_1 \sim 10^{19}$ Hz and $\nu_2 \sim 3 \times 10^{23}$ Hz.
The break at $\nu_1$ corresponds to the IC scattering of IR photons
by electrons with $\gamma \sim \gamma_\mathrm{br}$.
On the other hand, the break at $\nu_2$ is caused by IC scattering
of optical-UV photons by electrons with $\gamma \sim \gamma_\mathrm{br}$.
We assumed various values of $p_u$ in Figure \ref{fig:sed-fit-tuv1ev}.
The spectral shape at $\nu \gtrsim \nu_2$
does not depend on the value of $p_u$,
because the values of $u_\mathrm{opt\mathchar`-uv}$ and 
$q_\mathrm{inj, \, 0}$ are adjusted to fit the flux of GeV $\gamma$-rays.
It is to be noted that the photons with $\nu \gtrsim \nu_2$ are produced 
by IC scattering of optical-UV photons by electrons 
with $\gamma \sim \gamma_\mathrm{max}$ 
and that the scattering occurs in the Klein-Nishina regime.

It is noted that the radio emission of our models exceed 
that of Sgr A$^*$ for $\nu < 10^{10}$ Hz, 
but the model emission should be compared with that of 1 pc region from 
Sgr A$^*$.  This is shown in Figure 4 by a dotted line.
Some parameter values yield excess emission at $\lesssim 10^9$ Hz.
This excess is, however, avoided if the magnetic field is weaker
and the difference in the magnetic field does not affect the GeV emission
spectrum.   In Figure \ref{fig:sed-fit-tuv1ev} we fixed $B = 10^{-4}$ G
to compare the GeV spectra of various models.

When the GeV spectrum is fitted with different values of $p_u$,
the soft photon energy density in the optical-UV band should be adjusted with $p_u$.
Model A6 is presented to show the effect of the soft photons in the optical-UV band.
The parameters of A6 is different from those of model A3 only in
$u_\mathrm{opt\mathchar`-uv}$. 
The HE emission of A6 is produced mainly by IC scattering of IR photons.
It is found that the soft photons in the optical-UV band are important to
account for emission at $\nu \gtrsim 10^{23}$ Hz.
The effect of IC scattering of optical-UV photons in the GeV band 
becomes apparent for 
$u_\mathrm{opt\mathchar`-uv} \gtrsim 10^4$ eV cm$^{-3}$.

The electron kinetic energy density, $u_\mathrm{kin}$, is different
from model to model.
In model A3, the value of $u_\mathrm{kin}$
is $\sim 4.7 \times 10^{-7}$ erg cm$^{-3}$.
The electron kinetic energy contained in the emission region 
is $\sim 2.0 \times 10^{48}$ erg
and the energy injection rate of electrons is $1.3 \times 10^{39}$ erg s$^{-1}$.
(Note that this value is inversely proportional to $t_\mathrm{esc}$.)
This injected energy is mostly possessed by electrons 
with $\gamma < \gamma_\mathrm{br}$,
and the electrons emitting the GeV $\gamma$-rays have only a fraction of 
the injected energy:
the electrons with $\gamma > \gamma_\mathrm{br}$ contribute to the energy density 
only $\sim 1$ \%, when $p_u = 2.7$.
That is, the energy injection rate $\lesssim 10^{37}$ erg s$^{-1}$
is used to emit GeV $\gamma$-rays.
This number is just consistent with our flare model,
if the radiation efficiency during flares is low
and most of the kinetic energy of electrons is transported to 
the HE emission region.
In our flare model, the emission efficiency is found to be $\sim 10$ \% 
by numerical calculations.  
The observed flare luminosity of $\sim 10^{36}$ erg s$^{-1}$ means that 
$\sim 10^{37}$ erg s$^{-1}$ is injected into the flare emission region,
and that the most of the energy is directed to the environment without emission.
If the duty cycle of the flares is 10 \% as observed, 
10 \% of $10^{37}$ erg s$^{-1}$, i.e., $10^{36}$ erg s$^{-1}$, 
is directed to the steady GeV emission.
Considering various uncertainties regarding the flare models and GeV emission
region, the energetics of the current model is acceptable.
It is suggestive that the luminosities of the flare emission 
and the steady GeV emission are both $\sim 10^{36}$ erg s$^{-1}$.

In Figure \ref{fig:sed-fit-p27}, SEDs are shown for $p_u = 2.7$ and
various values of soft photon parameters.
The parameter values are given in Table \ref{table:param-values-27}.
The optical-UV emission with $T_\mathrm{opt\mathchar`-uv} = 3$ eV is assumed for 
models B1 and B2.
When the value of $T_\mathrm{opt\mathchar`-uv}$ is larger, 
larger values of $u_\mathrm{opt\mathchar`-uv}$ (B1) 
or $q_\mathrm{inj, \, 0}$ (B2) are required.
We found numerically that $u_\mathrm{opt\mathchar`-uv}$ should be 
$\gtrsim 10^5$ eV cm$^{-3}$ for $T_\mathrm{opt\mathchar`-uv} \gtrsim 3$ eV 
to fit the observed GeV spectrum. 
Larger values of $q_\mathrm{inj, \, 0}$
results in a poor fit as shown by model B2 in Figure \ref{fig:sed-fit-p27}.

In Figure \ref{fig:mag-mez}, SEDs for $p_u = 2.6$ are compared with
the emission expected from the central $30''$ given in \cite{mez96}.
Our models are calculated with $B = 10^{-4}$ G (solid line) 
and $3 \times 10^{-4}$ G (dash-dotted line).
The soft photon energy density is larger than the magnetic energy density
and the effect of the difference in the magnetic field appears 
only in the synchrotron radiation.
Synchrotron emission exceeds the radio flux observed in the central $30''$
when $B > 10^{-4}$ G.
Our model thus sets constraint on the average strength of the magnetic field
in the central 1 pc region.

In Figure \ref{fig:sed-fit-tev}, we show a model with 
$\gamma_\mathrm{max} = 6 \times 10^7$ to see the possibility of TeV emission
by our model.
Here we assumed a smaller value of the electron injection rate
than for the GeV emission models, i.e.,
$q_\mathrm{inj, \,0} = 3 \times 10^{-13}$ s$^{-1}$ and this corresponds to
the energy injection rate of $9.8 \times 10^{37}$ erg s$^{-1}$.
This is much smaller than for model A3 (dashed line),
whose energy injection rate is $1.3 \times 10^{39}$ erg s$^{-1}$.
Since the maximum Lorentz factor of electrons is much larger than our flare model,
efficient acceleration of electrons must occur during the propagation from
the flare region to the HE emission region.
Alternatively, electrons from other sources such as pulsar wind 
nebulae are responsible for the TeV emission.
This kind of model was presented by \cite{ad04} and \cite{ha07}.

\section{DISCUSSION}  \label{sec:discussion}

We have demonstrated that the GeV $\gamma$-ray spectrum from Sgr A$^*$ 
obtained by {\it Fermi} is well fitted by IC scattering in a region 
with a radius of $10^{18}$ cm, 
when soft photons are supplied by stars and dust.
We assumed that the nonthermal electrons are supplied by flare
events near the central black hole, 
which are often observed in the NIR and X-ray bands.
To fit the {\it Fermi} data, we assumed electrons with $p_u \gtrsim 2.6$
and $\gamma_\mathrm{max} \gtrsim 10^5$.
The success of our HE emission model, in turn, supports our NIR/X-ray flare model.

We found that the magnetic field in the region within $\sim 1$ pc
from the central black hole is $\lesssim 10^{-4}$ G 
because synchrotron emission exceeds the observed radio flux in the central
1 pc region.
The value of $\sim 10^{-4}$ G is also consistent with another constraint
given by \cite{cja11}.

Because the distance that nonthermal electrons travel from the flare region
to the HE emission region is $\sim 10^{18}$ cm,
the emission by those electrons during transport is to be considered.
When electrons are close to a flare region,
they emit radiation mostly by synchrotron emission.
Because the soft photon energy density in the HE emission region is 
$\sim 5 \times 10^4$ eV cm$^{-3}$,
synchrotron radiation dominates over IC scattering in regions
with $B \gtrsim 10^{-3}$ G.
Since the magnetic field of the flare emission region is $\sim 20$ G
and that in the HE emission region is $\sim 10^{-4}$ G,
synchrotron emission dominates over IC scattering in a region
within $\sim 10^{17}$ cm from the central black hole,
if the electron density is constant and 
the magnetic field decreases as $B \propto r^{-1}$, 
where $r$ is the distance from the central black hole.
Then the volume of the synchrotron dominant region is smaller than
that of the HE emission region by a factor of $10^{-3}$.
When the electron density decreases more rapidly than $r^{-2}$,
the contribution of the central region to the synchrotron emission is 
significant.
That is, a large flux of observed GeV emission suggests that nonthermal 
electrons do not follow a wind-like flow.
On the other hand, if $B \propto r^{-1}$ is assumed as above,
the decrease of the electron kinetic energy by synchrotron cooling is small.
For example, if the Lorentz factor of escaping electrons from a flare region
at $r \sim 10^{13}$ cm is $\gamma = 10^5$, it decreases to $\sim 2 \times 10^4$
at $r \sim 10^{17}$ cm and is almost constant for $r \gtrsim 10^{17}$ cm.
Here we assumed that electrons propagate at speed of light.
If the magnetic field decreases more rapidly than $r^{-1}$, 
this decrease in $\gamma$ becomes slower.
Therefore, the effect of radiative cooling during electron transport
from flare regions to the HE emission region is not important.

We assumed that the size of the HE emission region, $r_\gamma$,
is $10^{18}$ cm in this work.
Observationally there is no strong constraint on the value of $r_\gamma$,
except that the HE emission is consistent with no time variation 
during {\it Fermi} observation.
One possible constraint is that $r_\gamma \lesssim c t_\mathrm{IC}$ to fill
the region of $r_\gamma$ with electrons with $\gamma$ up to $\gamma_\mathrm{max}$,
where $t_\mathrm{IC}$ is the IC cooling time.
This sets the upper limit of 
$r_\gamma \lesssim 10^{20} (\gamma/10^5)^{-1}$ cm for 
$u_\mathrm{soft} \sim 5 \times 10^4$ erg s$^{-1}$.
On the other hand, the GeV emission by IC scattering in a more compact region 
near the central black hole is unlikely as follows.
In a region near the black hole, the soft photon source is
most likely the RIAF and the magnetic filed is approximately $\sim 1$ G.
When the GeV emission is by IC scattering off soft photons, 
there is a constraint such that $u_B < u_\mathrm{soft}$,
where $u_B$ is the energy density of the magnetic field.
This results in
\begin{equation}
    r_\gamma \lesssim \frac{1}{B} 
    \left( \frac{\epsilon L_\mathrm{bol}}{c} \right)^{1/2}
    \sim 6 \times 10^{12} \epsilon^{1/2} \left(\frac{1 \mathrm{G}}{B}\right)
    \left( \frac{L_\mathrm{bol}}{10^{36} \mathrm{erg} \,\, \mathrm{s}^{-1}}
    \right)^{1/2} \,\, \mathrm{cm},
\end{equation}
where $\epsilon$ is the scattered energy fraction of RIAF luminosity.
When $B = 1$ G, $\epsilon < 1$, and 
$L_\mathrm{bol} = 10^{36}$ erg s$^{-1}$, we obtain $r_\gamma < 10^{13}$ cm,
which size is comparable with a flare emission region of 
our leptonic flare model.
Then the HE emission region may exhibit time variation with timescale of
$r_\gamma/c \sim 300$ s, contrary to the observed steady GeV emission.

\begin{figure}
    \includegraphics[angle=0,scale=.65]{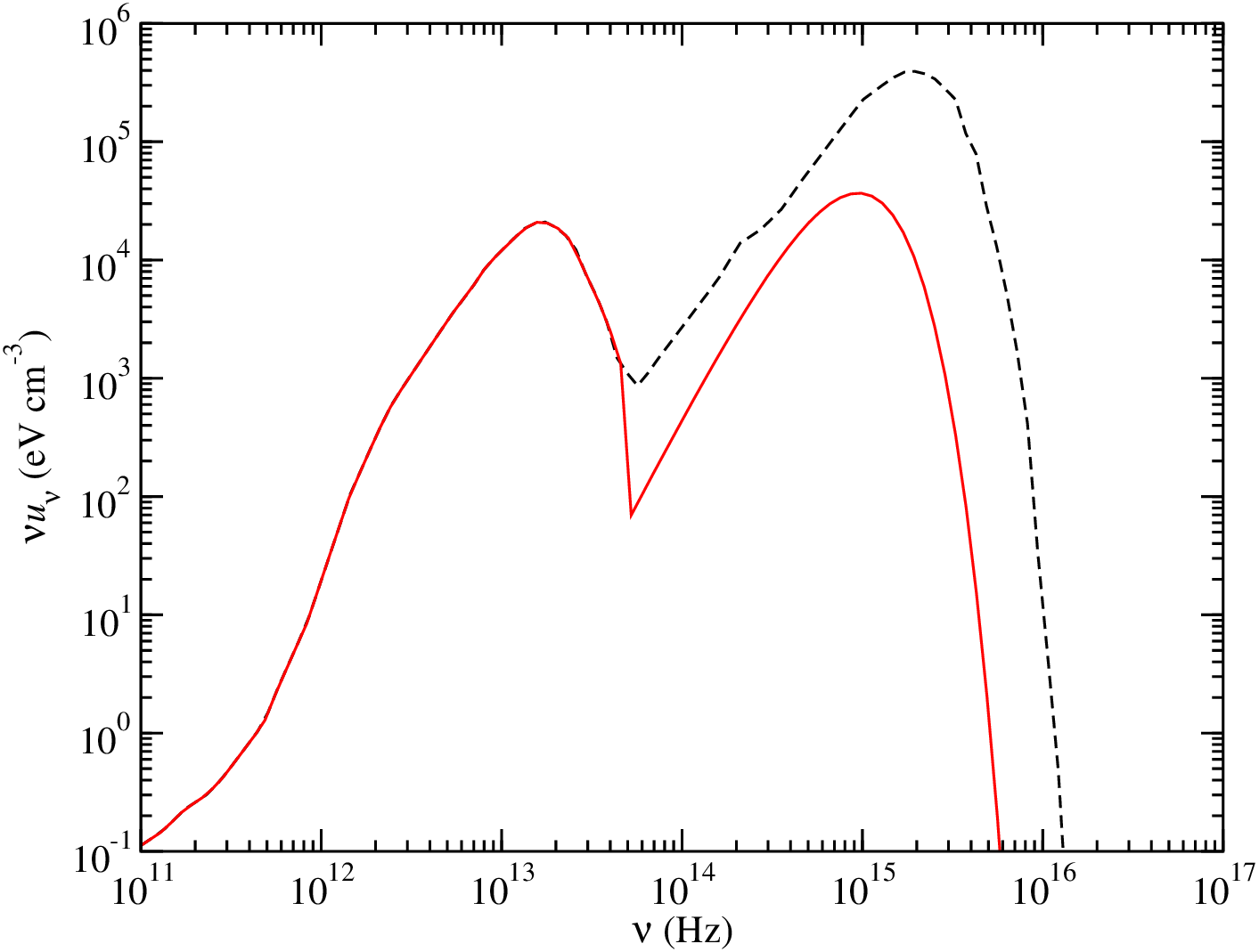}
    \caption{
      Soft photon energy density spectrum in the central region.
      The dashed line is adopted from \cite{mez96}.
      Our model (solid curve) uses their IR emission spectrum
      but the optical-UV spectrum is calculated as a blackbody 
      with parameters
      $T_\mathrm{opt\mathchar`-uv}$ and $u_\mathrm{opt\mathchar`-uv}$.
      The solid curve is calculated for 
      $T_\mathrm{opt\mathchar`-uv} = 1$ eV and 
      $u_\mathrm{opt\mathchar`-opt} = 5 \times 10^4$ eV cm$^{-3}$.
    }
    \label{fig:softphoton-spec}
\end{figure}

\begin{figure}

\includegraphics[angle=0,scale=.65]{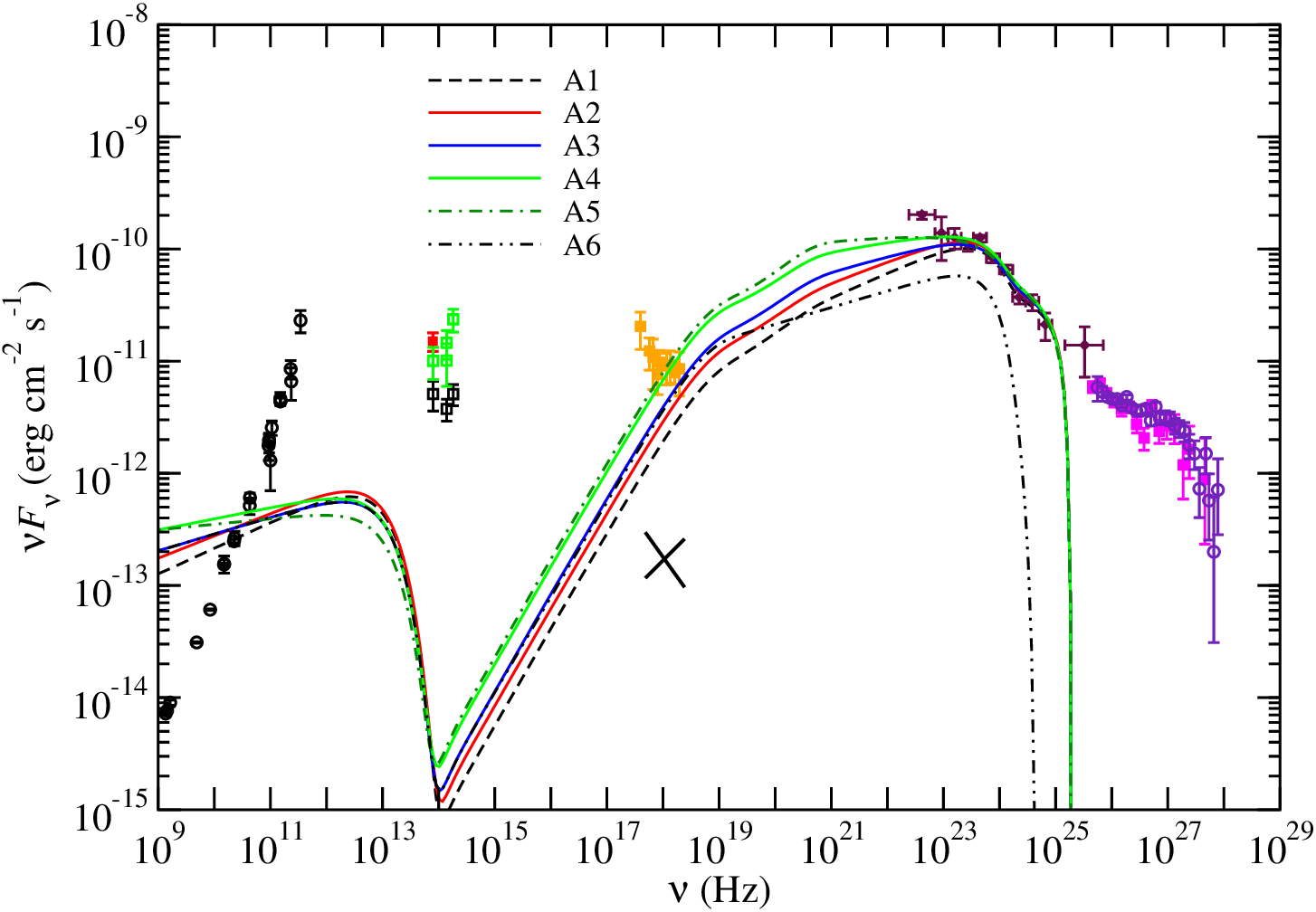}
\caption{
  SEDs for various values of $p_u$.  Here $\gamma_\mathrm{max} = 1.7 \times 10^5$,
  $B = 10^{-4}$ G, and $T_\mathrm{opt\mathchar`-uv} = 1$ eV.
  The value of $u_\mathrm{opt\mathchar`-uv}$ is changed to fit the data.
  Model parameters are given in Table \ref{table:param-values1}.
  The data in the range $10^{22}$ to $3 \times 10^{25}$ Hz are 
  obtained by {\it Fermi} \citep{cher11}.
  TeV emission data are from \cite{aha06} (filled squares) 
  and \cite{aha09} (open circles).
  Radio to submillimeter measurements are for the quiescent state
  \citep{mar01,zhao03} (open circles).
  IR data in the quiescent state are from \cite{gen03}.
  The X-ray data in the quiescent state are from \cite{bag03}.
  The flaring state in NIR (filled square) is taken from \cite{de09}.
  The X-ray flare data (filled squares) are from \cite{por08}.  
  \label{fig:sed-fit-tuv1ev}}
\end{figure}

\begin{figure}
\includegraphics[angle=0,scale=.65]{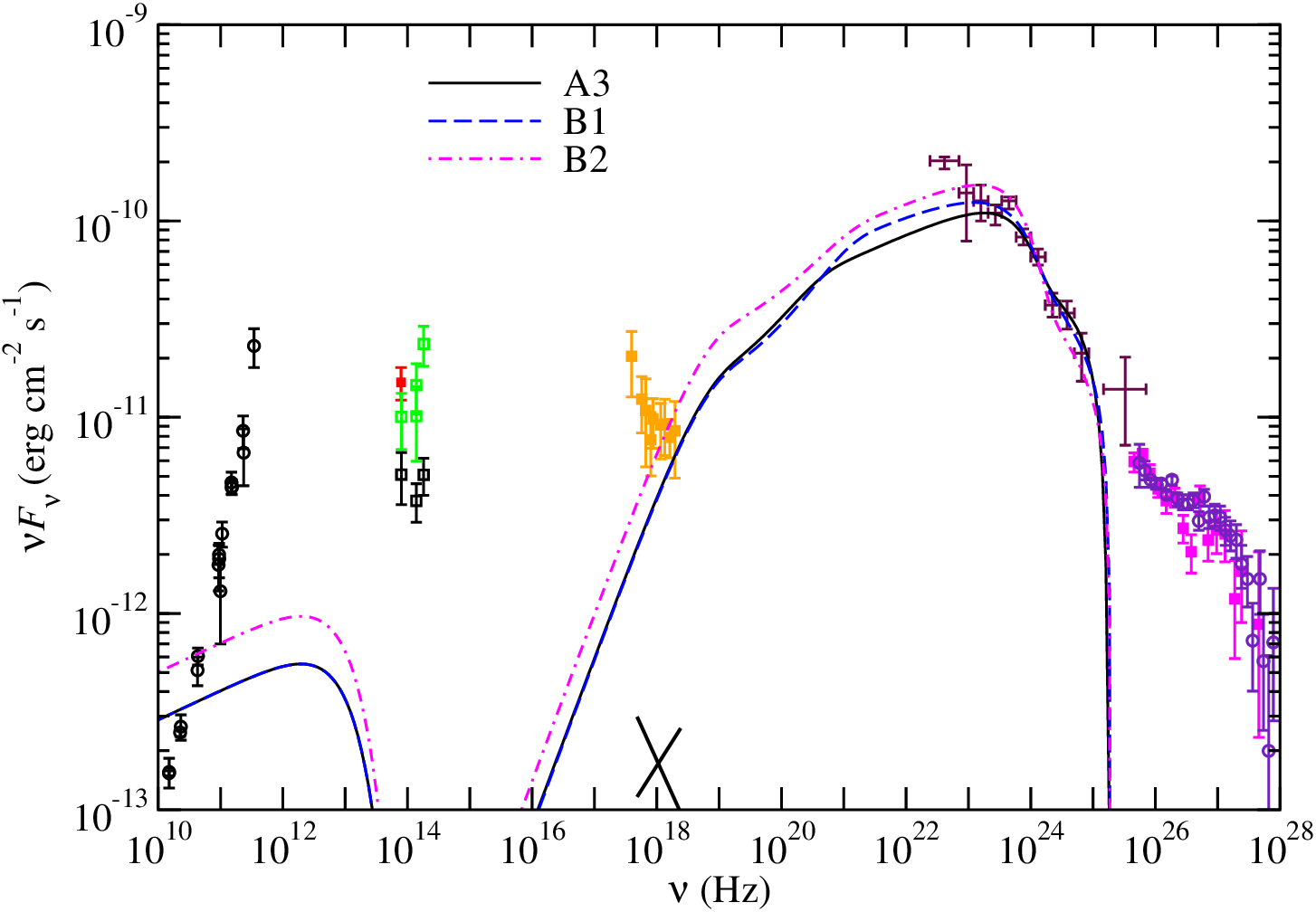}
\caption{
  SEDs with $p_u = 2.7$ for different soft photon parameters.
  The parameter values are given in Table \ref{table:param-values-27}.
  \label{fig:sed-fit-p27}}
\end{figure}

\begin{figure}
\includegraphics[angle=0,scale=.65]{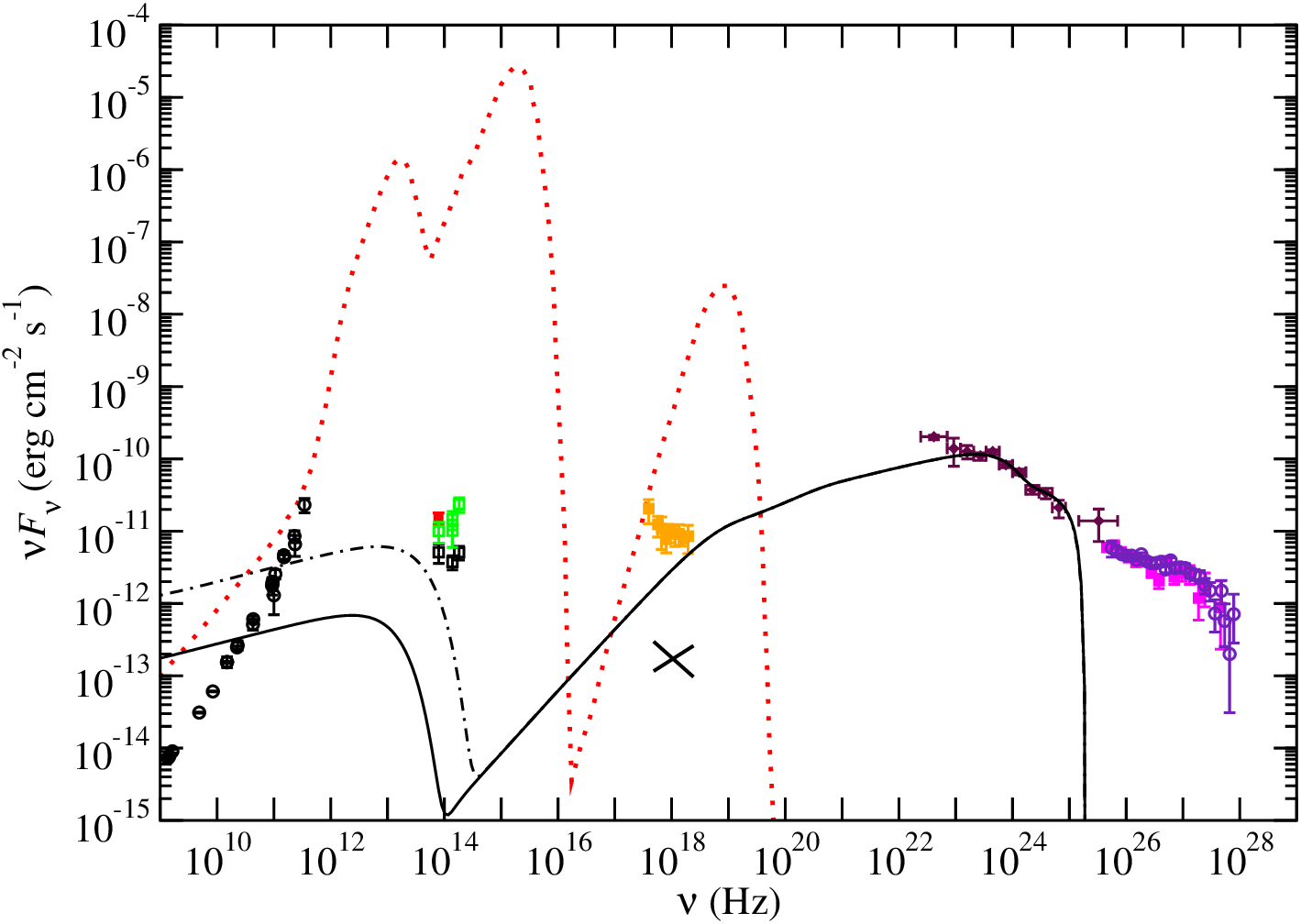}
\caption{
  Effect of magnetic field.
  Model SEDs with $B = 10^{-4}$ G (solid) and $B = 3 \times 10^{-4}$ G
  (dash-dotted) are compared with the emission (dotted) 
  in the central region $\lesssim 1.2$ pc \citep{mez96}.
  Here, $p_u = 2.6$, $\gamma_\mathrm{max} = 1.7 \times 10^5$, 
  $T_\mathrm{opt\mathchar`-uv} = 1$ eV, and
  $u_\mathrm{opt\mathchar`-uv} = 4 \times 10^4$ eV cm$^{-3}$ are assumed
  to calculate our models.
  \label{fig:mag-mez}
  }
\end{figure}

\begin{figure}
\includegraphics[angle=0,scale=.65]{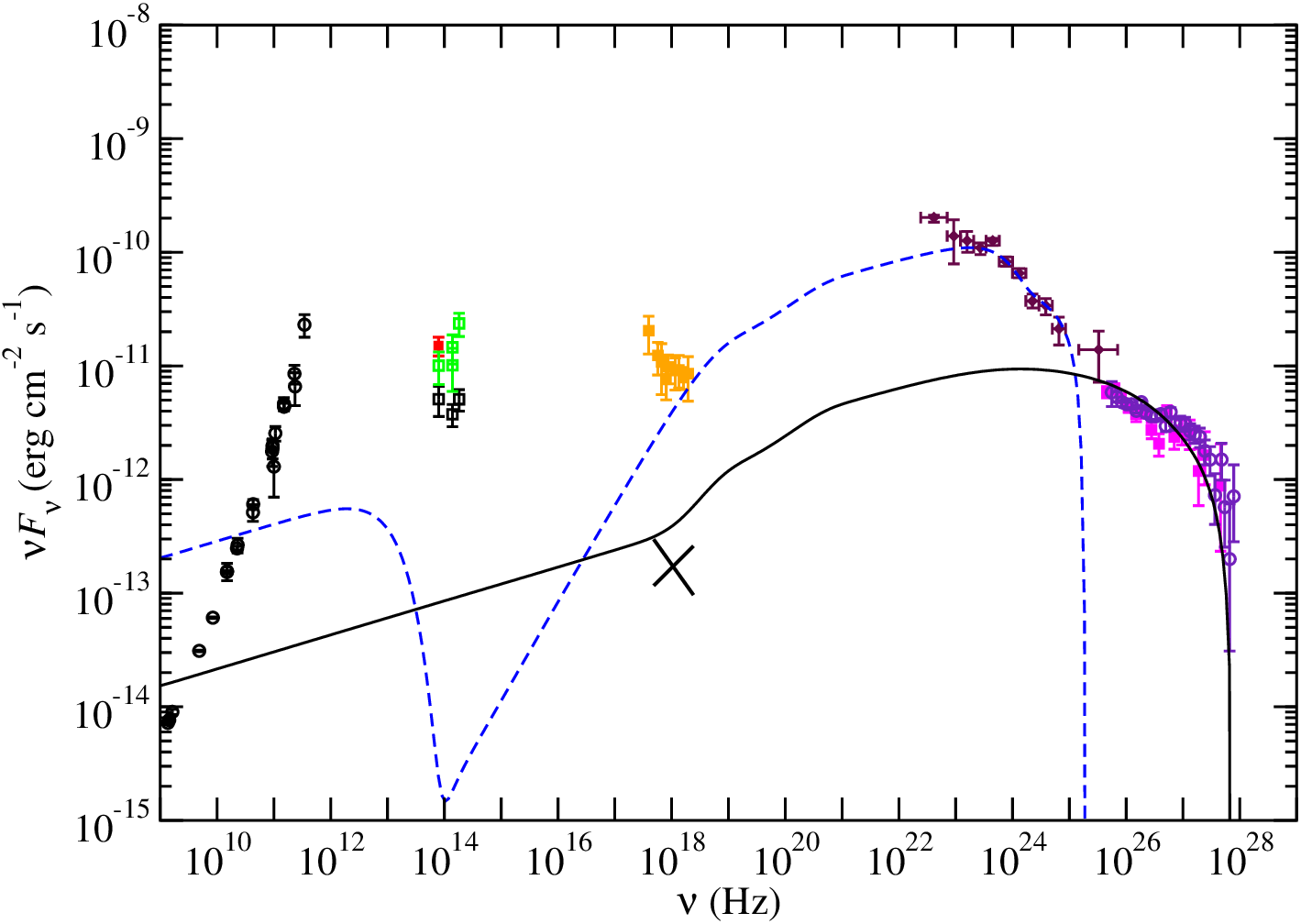}
\caption{
  Emission model of TeV $\gamma$-rays.
  The solid line is calculated for $\gamma_\mathrm{max} = 6 \times 10^7$.
  Here, $p_u = 2.7$, $B = 10^{-4}$ G, $T_\mathrm{opt\mathchar`-uv} = 1$ eV,
  and $u_\mathrm{opt\mathchar`-uv} = 5 \times 10^4$ eV cm$^{-3}$ are assumed.
  Model A3 with $\gamma_\mathrm{max} = 1.7 \times 10^5$ is shown by
  a dashed line for comparison.
  \label{fig:sed-fit-tev}
 }
\end{figure}

Because TeV $\gamma$-rays are absorbed by $e^+ e^-$ pair production in collisions
with soft photons, 
the soft photon density should be small enough to avoid the absorption
to account for the TeV emission observed by HESS.
The optical depth of the absorption is given by
$\tau_{\gamma \gamma} \sim  0.3 \sigma_\mathrm{T} r_\gamma n_s$
near the threshold, where $\sigma_\mathrm{T}$ is the Thomson cross section
and $n_s$ is the soft photon density.
Since $r_\gamma = 10^{18}$ cm, $n_s < 5 \times 10^6$ cm$^{-3}$ is required 
for $\tau_{\gamma \gamma} < 1$.
As shown in Figure \ref{fig:softphoton-spec}, the soft photon 
spectrum has two peaks at IR and optical-UV bands.
The photon density at the IR peak of $\nu \sim 2 \times 10^{13}$ Hz is
$\sim 2 \times 10^{5}$ cm$^{-3}$ and this gives 
$\tau_{\gamma \gamma} \sim 4 \times 10^{-2}$.
At $\nu \sim 10^{15}$ Hz, on the other hand, $n_s \sim 9 \times 10^{3}$ cm$^{-3}$ for 
$T_\mathrm{opt\mathchar`-uv} = 1$ eV and $u_\mathrm{opt\mathchar`-uv} = 5 \times 10^4$
erg cm$^{-3}$.  The optical depth is then $\tau_{\gamma \gamma} \sim 2 \times 10^{-3}$.
As mentioned in Section \ref{sec:results}, 
$T_\mathrm{opt\mathchar`-uv} \gtrsim 3$ eV needs $u_\mathrm{opt\mathchar`-uv} \gtrsim 10^5$
to fit the GeV emission.
Thus when the value of $T_\mathrm{opt\mathchar`-uv}$ is larger,
the central region becomes opaque for TeV $\gamma$-rays, 
contrary to the observations of TeV $\gamma$-rays.

\acknowledgments
This work has been partially supported by KAKENHI (F.T.: 20540231) from 
the Ministry of Education, Culture, Sports, Science and Technology of Japan.


\clearpage

\clearpage
\begin{deluxetable}{ccccccccc}
\tablecaption{Parameters
\label{table:param-values1}}  
\tablewidth{0pt}
\tablehead
{
\colhead{Model}  & \colhead{$p_u$} 
& \colhead{$u_\mathrm{opt\mathchar`-uv}$}
& \colhead{$q_\mathrm{inj, \, 0}$} \\
\colhead{} & \colhead{} & \colhead{(eV cm$^{-3}$)} & 
\colhead{(cm$^{-3}$ s$^{-1}$)}
}
\startdata
A1  & 2.54 & $4 \times 10^4$ & $2 \times 10^{-12}$ \\
A2  & 2.6  & $4 \times 10^4$ & $3 \times 10^{-12}$ \\ 
A3  & 2.7  & $5 \times 10^4$ & $4 \times 10^{-12}$ \\
A4  & 2.8  & $5 \times 10^4$ & $7 \times 10^{-12}$ \\ %
A5  & 2.9  & $7 \times 10^4$ & $8 \times 10^{-12}$ \\ %
A6  & 2.7  & $1$ & $4 \times 10^{-12}$ \\ 
\enddata

\tablecomments{All models assume $p = 1.3$, $\gamma_\mathrm{min} = 2$,
$\gamma_\mathrm{max} = 1.7 \times 10^{5}$, 
$r_\gamma = 10^{18}$ cm, $B = 10^{-4}$ G, and $T_\mathrm{opt\mathchar`-uv} = 1$ eV.
}
\end{deluxetable}

\clearpage
\begin{deluxetable}{ccccccccc}
\tablecaption{Parameters
\label{table:param-values-27}}
\tablewidth{0pt}
\tablehead
{
\colhead{Model}  
& \colhead{$T_\mathrm{opt\mathchar`-uv}$}
& \colhead{$u_\mathrm{opt\mathchar`-uv}$}
& \colhead{$q_\mathrm{inj,\, 0}$} \\
\colhead{} & \colhead{(eV)} &
\colhead{(eV cm$^{-3}$)} & \colhead{(cm$^{-3}$ s$^{-1}$)}
}
\startdata
B1  & 3  & $9 \times 10^4$ & $4 \times 10^{-12}$ \\ %
B2  & 3  & $4 \times 10^4$ & $7 \times 10^{-12}$ \\ %

\enddata

\tablecomments{All models assume $p = 1.3$, $\gamma_\mathrm{min} = 2$,
$\gamma_\mathrm{max} = 1.7 \times 10^{5}$, 
$r_\gamma = 10^{18}$ cm, and $B = 10^{-4}$ G.
}
\end{deluxetable}

\end{document}